\documentclass{aa}
\usepackage{graphicx}
\usepackage{txfonts}
\begin{document}
\title{Three active M dwarfs within 8~pc: 
L~449-1, L~43-72, \& LP~949-15 
\thanks{Based on our systematic search in archival data from 2MASS and SuperCOSMOS Sky Surveys, 
and spectroscopic observations with the ESO NTT (ESO 072.D-0295) }}

   \titlerunning{Three active M dwarfs within 8~pc} 


   \author{R.-D. Scholz
          \inst{1}
          \and
          G. Lo Curto
          \inst{2}
          \and
          R. A. M\'{e}ndez
          \inst{3}
          \and 
          V. Hambaryan
          \inst{1}
          \and
          E. Costa
          \inst{3}
          \and
          T. J. Henry
          \inst{4}
          \and
          A. D. Schwope
          \inst{1}
          }

   \offprints{R.-D. Scholz}

   \institute{Astrophysikalisches Institut Potsdam, An der Sternwarte 16,
              14482 Potsdam, Germany \\
              \email{rdscholz@aip.de, vhambaryan@aip.de, aschwope@aip.de}
            \and 
              European Southern Observatory,
              Alonso de Cordova 3107, Casilla 19001, Santiago 19, Chile \\
              \email{glocurto@eso.org}
            \and 
              Departamento de Astronomía, Universidad de Chile, 
              Casilla 36-D, Santiago, Chile \\
              \email{rmendez@das.uchile.cl, costa@das.uchile.cl}
            \and 
              Department of Physics and Astronomy, Georgia State University,
              Atlanta, GA 30302-4106 \\
              \email{thenry@chara.gsu.edu}
             }

   \date{Received ...; accepted ...}

\abstract{Three nearby star candidates were selected as bright 2MASS point 
sources without associated optical sources, i.e. with potentially large 
proper motions, subsequently confirmed by multi-epoch optical data from the 
SuperCOSMOS Sky Surveys. All three objects are listed in the NLTT catalogue 
of high proper motion stars. Follow-up spectroscopic observations allowed us 
to classify L~449-1 as M4.0e dwarf at 5.7~pc, L43-72 as M4.5e dwarf at 7.2~pc, 
and LP~949-15 as M5.0e dwarf at 6.1~pc, using known absolute $JHK_\mathrm{s}$
magnitudes of dwarfs with corresponding spectral types, respectively. 
All three stars exhibit $H{\alpha}$ 
emission lines, and all three can be identified with bright X-ray sources.  
The available ROSAT light curves of two of the objects show typical flare-like
variability. Thus, all three stars are active, very close 
potential neighbours of the Sun, previously not mentioned as such, certainly 
deserving further attention. In particular, these are very promising 
candidates for trigonometric parallax programs. 

   \keywords{surveys -- 
             astrometry --
             stars: kinematics -- 
             stars: low-mass, brown dwarfs --
             solar neighbourhood --
             X-rays: stars 
               }
   }

   \maketitle


\section{Introduction}
The recently growing interest in getting the nearby stellar sample completed
arises mainly from the fact that the nearest stars are excellent targets in
search programs for extra-solar planets. In addition, the nearest stars of a 
given mass, spectral type, metallicity, age, etc. can be studied in most
details and serve as benchmark sources in stellar physics. 
Finally, the solar neighbourhood offers the chance to study the 
stellar content and Galactic star formation history in a complete 
volume-limited sample. 

Surprisingly, our knowledge of the solar neighbourhood is still lacking
many stellar systems. Even in the immediate neighbourhood, within a 10~pc
horizon, where one may think that all stars have been registered, 
the number of missing systems is estimated between 25\,\% 
(Reid et al.\ \cite{reid03a}) and more than 30\,\% (Henry  et 
al.\ \cite{henry97}).

Many efforts have been started in recent years to fill the gaps. New
sky surveys in the near-infrared, like the DEep Near-Infrared Survey
(DENIS; Epchtein et al.\ \cite{epchtein97}) and the Two-Micron All Sky 
Survey (2MASS; Cutri et al.\ \cite{cutri03}) and the completion of optical
multi-colour and multi-epoch data bases like the SuperCOSMOS Sky Surveys
(SSS; Hambly et al.\ \cite{hambly01}), and other archives provide powerful 
tools in the search for our unknown neighbours. The most spectacular 
discoveries, at approximate distances of only 4~pc, include some late-M 
dwarfs (Delfosse et al.\ \cite{delfosse01}; Teegarden
et al.\ \cite{teegarden03}; Hambly et al.\ \cite{hambly04};
Henry et al.\ \cite{henry04})
as well as the brown dwarf binary $\varepsilon$\,Indi\,Ba,Bb 
(Scholz et al.~\cite{scholz03}; McCaughrean et al.\ \cite{mjm04}).
 
High proper motion catalogues, such as the New Luyten Two Tenths
(NLTT) catalogue (Luyten \cite{luyten7980}) do also still play an 
important role in the finding of new nearby stars, especially if combined 
with new accurate photometry, e.g. from 2MASS and spectroscopy (Scholz,
Meusinger \& Jahrei{\ss}~\cite{scholz01}; Reid \& Cruz \cite{reid02a};
Cruz \& Reid \cite{cruz02}; Reid, Kilkenny \& Cruz \cite{reid02b};
McCaughrean, Scholz \& Lodieu~\cite{mjm02}; Reid et al.\ \cite{reid03b};
Scholz, Meusinger \& Jahrei{\ss}~\cite{scholz05}).

In this short paper we present further three NLTT stars as newly
recognised close neighbours. \S~\ref{pmphot} describes the
selection of the candidates. \S~\ref{specobs} presents optical
classification spectroscopy and corresponding distance determination.
\S~\ref{xdata} summarises the available X-ray data; and in
\S~\ref{concdisc} we briefly discuss the results.

\section{Selection of nearby candidates}
\label{pmphot}

Nearby candidates in the southern sky were first detected as proper 
motion objects from comparing four different epoch observations in the SSS 
at the positions of bright ($J<14$) 2MASS point sources without optical 
identification. The initial sample of candidates was the same as described in
Scholz, Lodieu \& McCaughrean (\cite{scholz04}), i.e. $\sim$11000 objects
outside the Galactic plane.
There is a high probability that an object without optical 
counterpart (within 5~arcsec) in the 2MASS data base has a large proper 
motion ($>$0.3~arcsec/yr), since the epoch difference between 2MASS and 
the USNO A2.0 catalogue (Monet et al.\ \cite{monet98}) used for the 
cross-identification is typically about 15 years at southern declinations.
If a candidate was confirmed as moving object during the visual inspection,
then its accurate proper motion was determined using all available epochs
from the SSS, 2MASS and DENIS. The three objects listed in Table~\ref{sss2m} 
were by far the brightest 2MASS sources identified in this way as proper 
motion objects, which had no previous distance estimates. Some SSS and DENIS
magnitudes given in Table~\ref{sss2m} represent mean values 
from different epochs. More details can be found in the notes
below the table.

A query in VizieR\footnote{http://vizier.u-strasbg.fr/} allowed us
to identify the three stars with entries in the NLTT and with bright X-ray
sources: L~449-1 = 1RXS J051723.3$-$352152, LP~949-15 = 1RXS J060452.1$-$343331, 
and L~43-72 = 1RXS J181116.4$-$785919 
(Luyten \cite{luyten7980}; Voges et al.\ \cite{voges99}), although we note 
a large positional error ($\sim$90~arcsec) in the NLTT for L~449-1.

\begin{table*}
 \footnotesize
 \caption[]{Astrometry and photometry from the 2MASS, SSS, and DENIS.
}
\label{sss2m}
 \begin{tabular}{lcccccccccc}
 \hline
\hline
Name & $\alpha, \delta$ & Epoch & $\mu_{\alpha}\cos{\delta}$ & $\mu_{\delta}$ & $B_J$ & $R$ & $I$ & $J$ & $H$ & $K_s$ \\
             & (J2000) & & \multispan{2}{\hfil mas/yr \hfil} & \multispan{2}{\hfil (SSS) \hfil} & (DENIS)$^d$ & \multispan{3}{\hfil (2MASS) \hfil} \\
 \hline
 
L~449-1$^a$  & 05 17 22.92 $-$35 21 54.5 & 1998.995 & $-214$\,$\pm$\,4 & $-$169\,$\pm$\,6 & 13.023 & 10.743 &  8.897 &  7.400 &  6.854 &  6.558 \\ 
LP~949-15$^b$& 06 04 52.15 $-$34 33 36.0 & 1999.753 &  $+18$\,$\pm$\,4 & $+$335\,$\pm$\,2 & 13.784 & 11.579 &  9.027 &  7.742 &  7.183 &  6.866 \\ %
L~43-72$^c$  & 18 11 15.28 $-$78 59 22.7 & 2000.378 &  $+83$\,$\pm$\,7 & $+$308\,$\pm$\,9 & 13.736 & 11.504 &  9.490 &  7.840 &  7.327 &  6.964 \\ 
 \hline
 \end{tabular}

\scriptsize
{\bf Notes:}
Positions are from 2MASS, which provided accurate astrometry at the most 
recent epochs. Further details are given below. \\
$^a$ -- $\mu$ from 1 2MASS, 1 DENIS, and 4 SSS positions; Two SSS $R$
magnitudes, OR=10.697 and RE=10.789, were averaged; DENIS $J=7.262, K=6.484$. \\
$^b$ -- $\mu$ from 1 2MASS, and 4 SSS positions; Two SSS $R$ 
magnitudes, OR=11.702 and RE=11.455, were averaged. \\
$^c$ -- $\mu$ from 1 2MASS, 3 DENIS, and 5 SSS positions; Mean DENIS $J=7.811, 
K=7.023$; $R$ magnitude = RE, two other strongly deviating SSS $R$ magnitudes, OR=15.348 (?) and OR=16.470 (?),
were not used. \\
$^d$ -- For L~43-72, the mean of three independent, very similar measurements is 
given. In case of LP~949-15 DENIS data were not available so that the SSS $I$ 
magnitude is listed instead. The SSS $I$ magnitudes of L~449-1 and L~43-72 are
8.643 and 9.635, respectively.

\end{table*}
 
\normalsize

\section{Spectroscopy, distances, and kinematics}
\label{specobs}

Spectroscopic observations of L~449-1,  L~43-72, and LP~949-15, and
of two comparison stars, Gl~65B (M6.0) and Gl~213 (M4.0), were carried out 
with the EMMI instrument at the 3.5m New Technology Telescope (NTT), ESO
La Silla, on October 4th, 2003. We used grism\#4 and a 1~arcsec slit yielding
a spectral dispersion of 1.76~\AA/pix. With the $2\times2$ binning used, this 
corresponds to a resolution of about 600. The total exposure times for these 
very bright stars were less than 1 second. 

\begin{figure}[t] \resizebox{\hsize}{!}
{\includegraphics[bbllx=60,bblly=40,bburx=567,bbury=779,angle=0,clip=]
{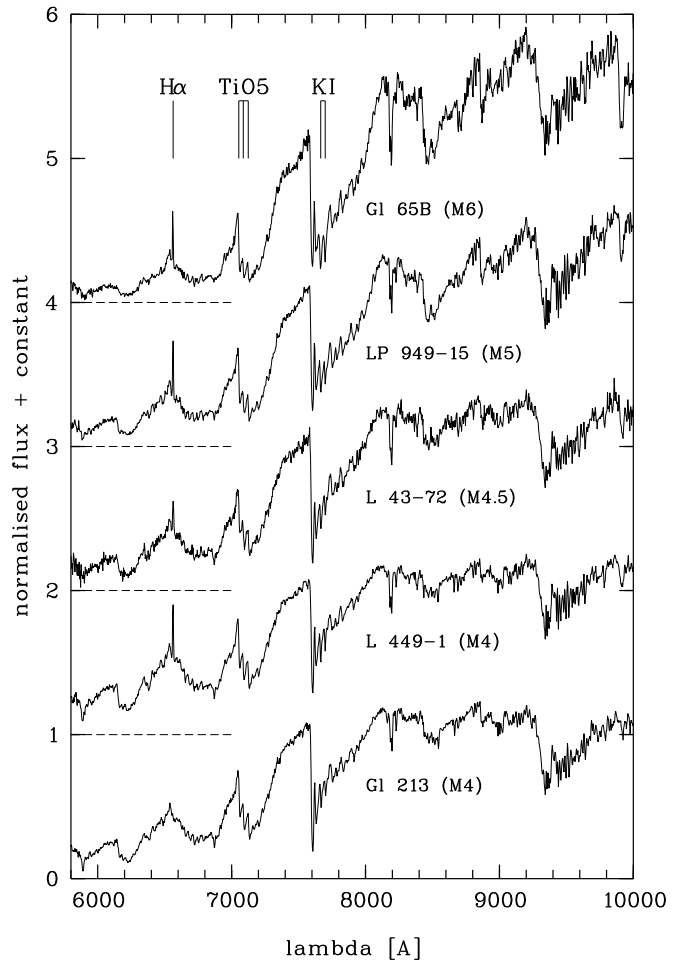}}
\caption{Optical spectra taken with EMMI on the ESO NTT. The two comparison
star spectra, obtained in the same run, are shown on top and at the bottom. 
Features used for a rough measurement of the radial velocities are labeled.
}\label{5spec}
\end{figure}

The reduction of the raw spectra was carried out using standard
ESO MIDAS routines, including an optimum extraction, cosmics removal, 
and extinction correction. Fluxes were calibrated using standard stars 
(LTT~7379 and LTT~2415) measured close in time to the objects. 
The fluxes of L~449-1 and LP~949-15 are better determined than the one for 
L~43-72 since the former were observed at the end of the night, when weather 
conditions were much more stable. L~43-72 was observed at relatively high air 
mass (1.65) due to its position on the sky. The fully calibrated
spectra normalised at 7500~\AA\, are shown in Fig.~\ref{5spec}, with the
two well known comparison stars, Gl~65B and  Gl~213 on top and at the
bottom of the plot, respectively.

We measured the spectral indices TiO5 and PC3 and obtained spectral types
using the formulae given in Reid, Hawley \& Gizis~(\cite{reid95}) and
Mart\'{\i}n et al.\ (\cite{martin99}), respectively (Table~\ref{spind}). 
The additional classification from the visual comparison with the spectra of
Gl~65B and  Gl~213 is also given in the table. The finally adopted spectral
type corresponds to the mean of the latter three estimates, rounded to half
a spectral sub-type.  The visual comparison clearly indicates a
later spectral type for LP~949-15 than for L~43-72, as can also be
concluded from the TiO5 measurements. 

\begin{table*}
 \footnotesize
 \caption[]{Spectral indices and types, radial velocities, distances, and
heliocentric space motions. Radial velocity errors are about 30~km/s.
}
\label{spind}
 \begin{tabular}{lcccccccccccc}
 \hline
\hline
Name & TiO5 & PC3 & SpT$_{\rm TiO5}$ & SpT$_{\rm PC3}$ & SpT$_{\rm comp}$ & SpT$_{\rm final}$ & $d_{{\rm spec}+JHK_\mathrm{s}}$ & $v_{\rm tan}$  & $v_{\rm rad}$  & $U$    & $V$    & $W$ \\
     &      &     &                  &                 &                  &                   & [pc]           & [km/s]         & [km/s]         & [km/s] & [km/s] & [km/s] \\        
 \hline
 
L~449-1  & 0.422 & 1.081 & M3.7 & M3.6 & M4.0 & M4.0 & 5.7$\pm$1.2 &  7 & $+$13 & $-1 \pm13$ & $-8 \pm22$ & $-13\pm17$ \\
LP~949-15& 0.301 & 1.171 & M5.0 & M4.3 & M5.0 & M5.0 & 6.1$\pm$1.3 & 10 & $+$67 & $-38\pm13$ & $-51\pm24$ & $-24\pm12$ \\ 
L~43-72  & 0.393 & 1.204 & M4.0 & M4.5 & M4.5 & M4.5 & 7.2$\pm$1.5 & 11 & $+$66 & $+51\pm19$ & $-37\pm19$ & $-23\pm13$ \\
 \hline
 \end{tabular}

\scriptsize
{\bf Notes:}
For L~449-1 and L~43-72 we can use the DENIS $IJK$ photometry
(Table~\ref{sss2m}) for an alternative distance determination
using the absolute magnitudes of M4.0 and M4.5 dwarfs given in
Kirkpatrick \& McCarthy (\cite{kirkpatrick94}). The results
of 5.5~pc (L~449-1) and 7.2~pc (L~43-72) are almost in exact
agreement with those obtained from the 2MASS photometry.
Using the absolute magnitudes
$M_J=8.37, 8.85, 9.30$, derived by Scholz,
Meusinger \& Jahrei{\ss} (\cite{scholz05})
for M4.0, M4.5, and M5.0 dwarfs, respectively,
we obtain distances of 6.4~pc, 6.3~pc, and 4.9~pc, 
for L~449-1, L~43-72, and LP~949-15, respectively.
The combination of the photographic SSS and 2MASS magnitudes
(Hambly et al.\ \cite{hambly04})
yields 6.5~pc, 6.1~pc, and 6.7~pc for L~449-1, LP~949-15,
and L~43-72, respectively.
 
\end{table*}
 
\normalsize

%

Distances were estimated using the 2MASS $JHK_\mathrm{s}$ magnitudes
from Table~\ref{sss2m} and mean absolute magnitudes of M4.0 to M5.0 dwarfs
from Kirkpatrick \& McCarthy~(\cite{kirkpatrick94}), conservatively
assuming uncertainties of 0.6~mag in the absolute magnitudes. 
This roughly corresponds 
to the maximum $M_J$ dispersion at spectral type M4.0, 
which is even larger than the $\sim$0.5~mag 
difference in $M_J$ between stars deviating
by half a spectral sub-class in the range of M3.5 to 
M5.5 (Scholz, Meusinger \& Jahrei{\ss} \cite{scholz05}).
All three stars fall in the 8~pc limit (Table~\ref{spind}).
The results of alternative distance estimates, given
in the Notes to Table~\ref{spind}, do not deviate
significantly from those listed in the table.

Rough estimates of the radial velocities were obtained by measuring the 
shift of the $H\alpha$ emission line, the KI doublet, and sharp TiO5 features
(marked in Fig.~\ref{5spec}) with respect to the two comparison spectra 
taken during the same night. The radial velocities of the comparison stars 
Gl~213 and Gl65B are known as $+104.3$~km/s and $+29.0$~km/s, respectively 
(Barbier-Brossat \& Figon \cite{barbier00}). 
The relative measurements were corrected for the Earth's
motion. The resulting mean radial velocities with an accuracy
of about 30~km/s are listed in Table~\ref{spind}.

Combining all the available information we computed heliocentric space 
motions following Johnson \& Soderblom~(\cite{johnson87}). The $UVW$ velocity
components of all three stars are not very significant due to the relatively
large errors. However, such small motions are typical of the Galactic thin 
disk population.

\section{X-ray data}
\label{xdata}

\begin{figure}[t] \resizebox{\hsize}{!}
{\includegraphics[bbllx=77,bblly=370,bburx=541,bbury=718,angle=0,clip=]
{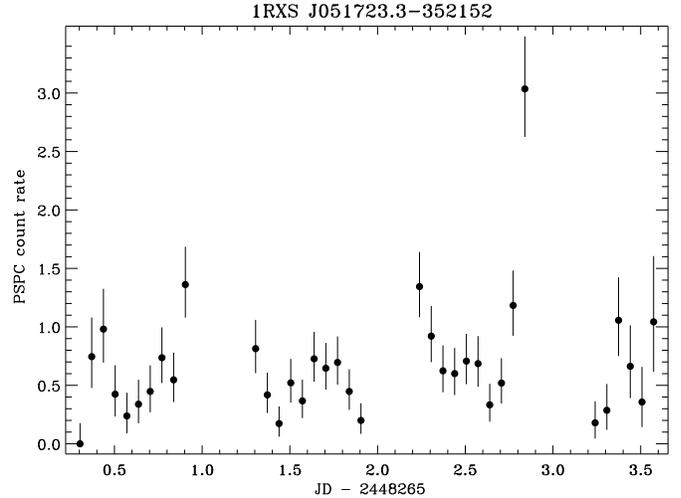}}
\caption{Light curve from ROSAT All Sky Survey observations of 1RXS~J051723.3$-$352152,
which we identified with L~449-1.
Shown are the background subtracted and vignetting
corrected count rates as a function of time. 
}\label{05_fig}
\end{figure}

\begin{figure}[t] \resizebox{\hsize}{!}
{\includegraphics[bbllx=77,bblly=370,bburx=541,bbury=718,angle=0,clip=]
{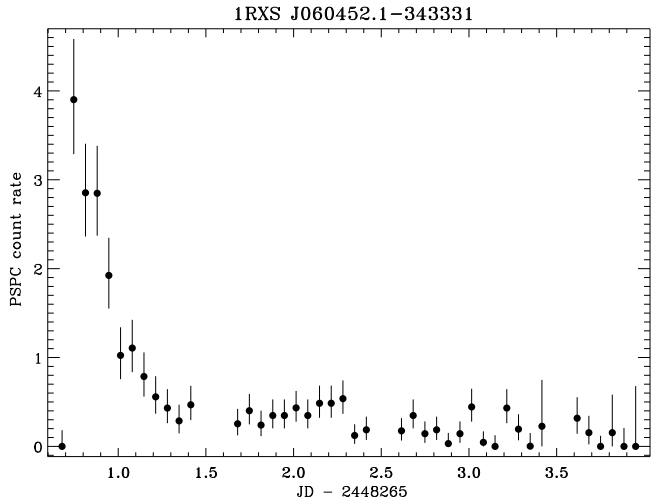}}
\caption{Same as in Fig.~\ref{05_fig}, but for 1RXS~J060452.1$-$343331
identified with LP~949-15.
}\label{06_fig}
\end{figure}

All three objects were detected as bright X-ray sources in the ROSAT
all-sky survey (Voges et al.\ \cite{voges99}), with mean count rates 
between 0.4 and 0.6 counts per second. Two of the sources
have long enough exposure times of approximately 600 seconds so that their
light curves can be plotted (Fig.~\ref{05_fig} and Fig.~\ref{06_fig}). These
two objects were already mentioned as variable X-ray sources by Fuhrmeister \&
Schmitt (\cite{fuhrmeister03}). Their light curves show very significant
variations, which can be interpreted as flares, perhaps with associated 
irregular X-ray brightness variations of comparatively lower amplitudes.

L~449-1 and LP~949-15 were already mentioned as dMe counterparts of 
ROSAT Wide Field Camera (WFC) sources by Pounds et al.\ (\cite{pounds93}).
X-ray variability was not detected in the ROSAT WFC data 
(McGale et al.\ \cite{mcgale95}), most likely due to the much poorer
count rate statistics.
All three objects are also listed as EUVE sources 
(e.g. Malina et al.\ \cite{malina94}; Bowyer et al.\ \cite{bowyer96}).
LP~949-15 appears in the QORG catalogue of radio/X-ray sources
by Flesch \& Hardcastle (\cite{flesch04}).

Our estimates of X-ray and bolometric luminosities yielded
$\log{(L_{X}/L_{bol})} \approx -4.0...-3.0 $ for the three stars,
where the lower and higher values correspond to the mean and maximum
level of the light curves, respectively. These results are in good
agreement with those for other M4.0-M5.0 dwarfs (see, for example,
Fig.~5 in Hambaryan et al. \cite{vvh}).

%

%

\section{Discussion}
\label{concdisc}

LP~949-15 is listed as AP~Col, a variable star of UV Cet type, in the 
General Catalog of Variable Stars (Samus et al.\ \cite{samus04}).
The amplitude of its variability in the $V$ band is given as 2.5~mag.
The available SSS photometry (two independent $R$ magnitudes of
11.70 and 11.45, respectively), the SSS colours (Table~\ref{sss2m},
and the additional photometry ($V=13.17$) from the Carlsberg Meridian 
Catalogs (CMC \cite{cmc99}) do not hint at strong variability. In addition,
our distance estimate based on the near-infrared photometry (6.1~pc),
is in reasonable good agreement with alternative estimates based on the 
CMC $V$ magnitude (7.0~pc) and SSS $I$ magnitude (4.8~pc), if we use
again absolute magnitudes of M5.0 dwarfs from Kirkpatrick \& McCarthy 
(\cite{kirkpatrick94}). 
Therefore, we think that our distance estimates for LP~949-15 are 
not strongly biased by its flare activity.
Even taking into account the relatively large
uncertainties in our distance estimates, all
three stars are very likely to be within 10~pc.

L~449-1 was previously thought to be a carbon star 
(Alksnis et al.\ \cite{alksnis01}), which is ruled out by our
findings. Other interesting cross-identifications include those
of L449-1 and L~43-72 with the IRAS faint source catalogue 
(Moshir et al.\ \cite{moshir89}).

Although all three stars are known X-ray sources and were earlier
classified as proper motion stars, we are not aware of any
publication treating them as nearby stars. Our cross-identification
with the 2MASS providing accurate photometry and the follow-up
low-resolution spectroscopy allowed us to uncover them as close
neighbours to the Sun. As such, they are certainly worth investigating
further, with more accurate distances to be obtained in a trigonometric
parallax program.


\begin{acknowledgements}
Our search for nearby star candidates was primarily based on data from 
the Two-Micron All Sky Survey, a joint project of the University of Massachusetts 
and the Infrared Processing and Analysis Center/California Institute of Technology, 
funded by the NASA and NSF, and from the SuperCOSMOS Sky Surveys at the Wide-Field 
Astronomy Unit of the Institute for Astronomy, University of Edinburgh. 
The spectroscopic observations were carried out with the ESO NTT. 
We acknowledge the use of the Simbad database  and
the VizieR Catalogue Service operated at the CDS. 
EC and RAM acknowledge support by the Fondo Nacional de
Investigaci\'on Cient\'{\i}fica y Tecnol\'ogica (proyecto No. 1010137,
Fondecyt) and by the Chilean Centro de Astrof\'{\i}sica FONDAP
(No. 15010003). We thank the referee, Neill Reid, for his prompt 
report and helpful comments. 

\end{acknowledgements}


\end{document}